\documentclass[12pt,twoside,english,refpage,intoc,bibliography=totoc,index=totoc,BCOR7.5mm,captions=tableheading,numbers=noenddot]{revtex4-1}
\usepackage[T1]{fontenc}
\usepackage[latin9]{inputenc}
\usepackage{color}
\usepackage{babel}
\makeatletter
\addto\extrasfrench{%
   \providecommand{\fg}{\ifdim\lastskip>\z@\unskip\fi~\frqq}%
}

\makeatother
\usepackage{textcomp}
\usepackage{amsmath}
\usepackage{amsthm}
\usepackage{amssymb}
\usepackage[unicode=true,
 bookmarks=true,bookmarksnumbered=true,bookmarksopen=false,
 breaklinks=false,pdfborder={0 0 1},backref=false,colorlinks=true]
 {hyperref}
\hypersetup{pdftitle={Guide de l'utilisateur de LyX},
 pdfauthor={L'équipe LyX, Traduction: Adrien Rebollo et Siegfried Meunier-Guttin-Cluzel},
 pdfsubject={LyX},
 pdfkeywords={LyX},
 linkcolor=black, citecolor=black, urlcolor=blue, filecolor=blue, pdfpagelayout=OneColumn, pdfnewwindow=true, pdfstartview=XYZ, plainpages=false}

\makeatletter
\numberwithin{equation}{section}
\numberwithin{figure}{section}
\usepackage{ifpdf} % part of the hyperref bundle
\ifpdf % if pdflatex is used

 \IfFileExists{lmodern.sty}{\usepackage{lmodern}}{}

\fi % end if pdflatex is used

\usepackage[figure]{hypcap}

\let\myTOC\tableofcontents
\renewcommand\tableofcontents{%
  \frontmatter
  \pdfbookmark[1]{\contentsname}{}
  \myTOC
  \mainmatter }

\usepackage{fancyhdr}
\@ifundefined{extrarowheight}
 {\usepackage{array}}{}
\makeatother

\begin{document}

\title{Equivalence between nonlinear dynamical systems and urn processes}

\author{Léon Brenig\textsuperscript{*}}
\address{\textsuperscript{1}Service de Physique des Systèmes Dynamiques,
Faculté des Sciences, Université Libre de Bruxelles, 1050 Brussels,
Belgium}
\email{lbrenig@ulb.ac.be}

\author{Iram Gleria}
\address{\textsuperscript{2}Instituto de Física, Universidade Federal de
Alagoas, 57072-970 Maceió, AL, Brazil}

\author{Tarcísio M. Rocha Filho}
\address{\textsuperscript{3}Instituto de Física and International Center
for Condensed Matter Physics, Universidade de Brasília, 70919-970
Brasília, DF, Brazil}

\author{Annibal Figueiredo}
\address{\textsuperscript{3}Instituto de Física and International Center
for Condensed Matter Physics, Universidade de Brasília, 70919-970
Brasília, DF, Brazil}

\author{Benito Hernández-Bermejo}
\address{\textsuperscript{4}Departamento de Biologia y Geologia, F\'\i sica y
Qu\'\i mica Inorg\'anica, Universidad Rey Juan Carlos. Calle Tulip\'an S/N,
28933-M\'ostoles-Madrid, Spain}

\begin{abstract}
An equivalence is shown between a large class of deterministic dynamical
systems and a class of stochastic processes, the balanced urn processes.
These dynamical systems are governed by quasi-polynomial differential
systems that are widely used in mathematical modeling while urn processes
are actively studied in combinatorics and probability theory. The
presented equivalence extends a theorem by Flajolet et al.\
(Flajolet, Dumas and Puyhaubert
Discr.\ Math.\ Theor.\ Comp.\ Sc.\ AG - 2006, DMTCS Proceedings)
already establishing an isomorphism between urn processes and a particular
class of differential systems with monomial vector fields. The present
result is based on the fact that such monomial differential systems
are canonical forms for more general dynamical systems. 
\end{abstract}
\maketitle

\section{Introduction}
\label{sec1}

In this work we establish a quite general equivalence between two
apparently unrelated fields of mathematics, nonlinear differential
dynamical systems and urn processes. As will be shown in the sequel,
to any urn process corresponds an infinite equivalence class of nonlinear
differential systems. The urn models are stochastic processes while
the differential dynamical systems are deterministic. Although
at first sight such a bridge between stochastic and deterministic
processes could seem unlikely, the clue, however, is that while the
time evolution of a stochastic process is random by definition, the
time evolution of its probability density is deterministic. As shown
in the sequel, the evolution equation for the probability density
of a balanced urn process is related to a wide class of differential
dynamical systems that are used in many fields of mathematical modeling.
These results opens the way to knowledge transfer between both fields
of research.

Urn processes have first been introduced by Laplace~\cite{key-2}
and later re-introduced and systematically studied for two-colours
models by Pólya~\cite{key-3}. Nowadays, they are the object of an
intense theoretical activity involving combinatorics and probability
theory~\cite{key-4}. They provide a powerful modeling tool in many
scientific domains such as statistical physics, population genetics,
epidemiology, economy and some social phenomena such as innovation
diffusion~\cite{key-5}. 

An urn process consists of three items: a box, the so-called urn,
containing objects that can differ by some distinctive features, an
infinite reservoir of such objects and a given set of replacement
rules of the objects in the urn. The paradigm of such a process is
an urn containing balls that only differ by their colours, with $N$
possible colours. The replacement rules are the prescriptions that
make the composition of the urn evolve at each discrete step. Thus,
at each step a ball is randomly picked from the urn, with equal chance
for all balls present in the urn. Its colour is noted and the ball
is reintroduced in the urn. Depending on the colour of the picked
ball, fixed numbers of balls of each colour are then transferred to
the urn from the reservoir. These integer numbers, $M_{ij}$, with $i,j\in\left\{ 1,...,N\right\} $
form a matrix, the replacement matrix. An entry $M_{ij}$ means that
if the ball drawn from the urn is of colour $i$, then one has to transfer
$M_{ij}$ balls of colour $j$ from the reservoir into the urn, for each $j$. Some
entries can be negative, in which case the balls of the corresponding
colours are transferred from the urn to the reservoir at each step.
Frequently, the entries are non-negative but if some of the diagonal
entries are negative, conditions must be imposed in order to avoid
blocking the process. The collection of the numbers of balls of each
colour in the urn at a given step $n$ forms its composition vector
at that step, $U_{n}$ and the sequence $(U_{n};n\geq0)$, where the
initial composition $U_{0}$ of the urn is given, represents the urn
process up to the $n$-th step.

The equivalence property presented in this work is limited to the
so-called balanced urn processes. For such processes, the total number
of balls replaced at each step in the urn is independent of the colour $i$
of the picked ball, that is, $\sum_{j=1}^{N}M_{ij}=\sigma$ for all
$i$, where $\sigma$ is called the balance. Flajolet and co-workers~\cite{key-1}
showed an isomorphism between balanced urn processes and certain systems
of autonomous ordinary differential equations with monomial vector
fields.

Differential dynamical systems are systems of autonomous first order
ordinary differential equations (ODEs) of the type 
\[
\frac{dx_{i}(t)}{dt}=f_{i}(x_{1}(t),...,x_{N}(t));i=1,...,n
\]
 where the $x_{i}$ are real functions of the time $t$ and the functions
$f_{i}$ can be quite general. The only constraint on the functions
$f_{i}$ are dictated by the fundamental laws of the phenomena to
be modeled and by the existence and unicity of the solutions to the
Cauchy problem. In most cases they are nonlinear functions. Such systems
are ubiquitous in the mathematical modeling in physics, chemistry,
biology, ecology, economical and social sciences~\cite{key-6,key-7,key-8}.
Generally, due to their nonlinearity, these systems are non-integrable
and present a rich diversity of behaviours. It is well known that
their solutions are very sensitive to the functional form of the nonlinear
functions $f_{i}$. Nevertheless, it has been shown that a large class
of such systems can be brought to two canonical forms. This is shown
in two steps.

The first step uses the property that many dynamical systems can be
brought to the quasi-polynomial, also called generalized Lotka-Volterra,
form~\cite{key-9,key-10}:
\[
\frac{du_{i}}{dt}=u_{i}\sum_{j=1}^{M}A_{ij}\prod_{k=1}^{N}u_{k}^{B_{jk}}
\]
with $i\in\{1,..,N\}$ and where we omitted the dependence in $t$
of the dependent real variables $u_{i}$. The matrices
$A$ and $B$ are any rectangular matrices with real and constant entries.

The second step rests on the fact, independently discovered at least
three times~\cite{key-11,key-12,key-13}, that
systems in the quasi-polynomial format and for which the variables
$u_{i}$ remain positive, can be transformed into two canonical forms.
One of these forms corresponds to the well-known Lotka-Volterra systems
commonly used in population dynamics. The other canonical form is
a system of ODEs with monomial functions of the dependent variables
in the right-hand-side that we called the monomial canonical form.
This type of differential systems is much less known and used in models.
It has, thus, been a surprise to us to discover a work~\cite{key-1} showing that it is related
to the balanced urn stochastic processes.

In the next Section, we briefly summarize the derivation of the two
canonical forms for quasi-polynomial systems. In Section~\ref{sec3}, we
recall the proof of the isomorphism between balanced urn processes
and the monomial differential systems. Section~\ref{sec4} is devoted to
the statement and proof of our main result and to some of its consequences.
Conclusions and perspectives are discussed in Section~\ref{sec5}.

\section{Canonical forms}
\label{sec2}

Let us consider the set of all dynamical systems that can be brought
to the quasi-polynomial (QP) form
\begin{equation}
\frac{dx_{i}}{dt}=x_{i}\sum_{j=1}^{N}A_{ij}\prod_{k=1}^{n}x_{k}^{B_{jk}},\qquad\mbox{ for }i=1,\ldots,n,
\label{eq:1}
\end{equation}
and such that all the functions $x_{i}(t)$ remain positive for all
$t$. The rectangular matrices $A$ and $B$ have real and constant
entries. $N$ and $n$ do not need to be equal. Let us call the above
form $QP(A,B)$.
Under the action of the following monomial transformations
\begin{equation}
x_{i}=\prod_{k=1}^{n}\tilde{x}_{k}^{C_{ik}}\qquad\mbox{ for }i=1,\ldots,n,
\label{eq:2}
\end{equation}
where $C$ is any real, constant and invertible $n\times n$ matrix,
the system~(\ref{eq:1}) becomes
\begin{equation}
\frac{{d\tilde{x}}_{i}}{dt}=\tilde{x}_{i}\sum_{j=1}^{N}\tilde{A}_{ij}\prod_{k=1}^{n}\tilde{x}_{k}^{\tilde{B}_{jk}}\qquad\mbox{ for }i=1,\ldots,n,
\label{eq:3}
\end{equation}
with
\begin{equation}
\tilde{A}=C^{-1}A,
\label{eq:4}
\end{equation}
and 
\begin{equation}
\tilde{B}=BC,
\label{eq:5}
\end{equation}
where the product is the matrix product. A transformation~(\ref{eq:2}), thus, sends the system $QP(A,B)$ to the system $QP(C^{-1}A, BC)$.\\ 
From equations~(\ref{eq:4})
and~(\ref{eq:5}) it is obvious that 
\begin{equation}
\tilde{B}\tilde{A}=BA\label{eq:6}
\end{equation}
so that the $N\times N$ matrix $M=BA$ is an invariant of transformations~(\ref{eq:2}).
In other words, all the $QP(A,B)$ having the same product matrix
$BA=M$ belong to a same equivalence class represented by this
matrix. Moreover, the transformations~(\ref{eq:2}) are diffeomorphisms
and form a group.

Now, let us take as new variables the $N$ monomials that appear in
the right-hand-side of equation~(\ref{eq:1})
\begin{equation}
u_{j}=\prod_{k=1}^{n}x_{k}^{B_{jk}},
\label{eq:7}
\end{equation}
for $j\in\{1,...,N\}$ and calculate the time-derivative of these
variables using equations~(\ref{eq:1}). This yields
\begin{equation}
\frac{du_{j}}{dt}=u_{j}\sum_{_{k=1}}^{N}M_{jk}u_{k},
\label{eq:8}
\end{equation}
with $j\in\{1,...,N\}$ and $M=BA$. This is the first canonical form,
the Lotka-Volterra (LV) systems already known from population dynamics.
In the latter field, the $N\times N$ matrix $M$ is named the interaction matrix of
the system. The meaning of the equivalence class related to the invariant matrix
$BA$ becomes now evident: the systems $QP(A,B)$ belonging to a given
equivalence class are all equivalent to the same LV system~(\ref{eq:8}) with the
interaction matrix $M$ equal to $BA$. Let us also notice that the above system corresponds 
to the $QP(M,I)$ system, where $I$ is the $N\times N$ identity matrix

Clearly, if $N>n$ the phase-space associated to the LV system~(\ref{eq:8})
is higher dimensional than the phase-space of the original QP system~(\ref{eq:1}).
This means that the variables $u_{j}$ are not all independent and
that the trajectories corresponding to the original system are confined
to invariant subspaces in the $N$-dimensional positive orthant of
the phase-space of equation~(\ref{eq:8}). Conversely, for $n>N$
the transformation to the LV system corresponds to a reduction of
the number of variables of the original $QP(A,B)$. More results on
the transformation to the LV canonical form can be found in references~\cite{key-14,key-15,key-16}.
Let us just remark that for $N\neq n$, the transformation~(\ref{eq:7})
is not bijective. In the case $N=n$, the transformation is a bijection
and belongs to the group of monomial transformations~(\ref{eq:2}).

The second canonical form is obtained in the following way. As
the canonical form~(\ref{eq:8}) is a $N$-dimensional QP system of the form $QP(M,I)$, we can apply on it 
a monomial transformation of type~(\ref{eq:2}) . Provided $M$ is invertible 
we can choose the form of the matrix $C$ as $C=M$ yielding:
\begin{equation}
u_{k}=\prod_{j=1}^{N}\tilde{u}_{j}^{M_{kj}},
\label{eq:9}
\end{equation}
Using relations~(\ref{eq:4}) and~(\ref{eq:5}) for $A=M$ and $B=I$
one, then, obtains
\begin{equation}
\frac{d\tilde{u_{j}}}{dt}=\tilde{u_{j}}\prod_{k=1}^{N}\tilde{u}_{k}^{M_{jk}},
\label{eq:10}
\end{equation}
with $j\in\{1,...,N\}$.

System~(\ref{eq:10}) represents the second canonical form of interest here. Again,
this means that all the $QP(A,B)$ systems such that their matrix product
$BA=M$ are equivalent to this canonical form which can be noted
$QP(I,M)$. 

Let us summarize the above results in the following theorem, with $\dot v\equiv dv(t)/dt$:

\textbf{Theorem 1:}\textit{ All the $QP(A,B)$ systems with the same
$N\times N$ product matrix $BA=M$ are equivalent to two canonical
forms, $\dot{u}_{j}=u_{j}\sum_{_{k=1}}^{N}M_{jk}u_{k}$, $j=1,...,N$,
and $\dot{\tilde{u_{j}}}=\tilde{u_{j}}\prod_{k=1}^{N}\tilde{u}_{k}^{M_{jk}}$,
$j=1,...,N$. The trajectories of all n-dimensional $QP(A,B)$ systems
with the same product matrix $BA$ are mapped one into the other by
the transformations $x_{i}=\prod_{k=1}^{n}\tilde{x}_{k}^{C_{ik}}$
, $i=1,...,n$ for any invertible real matrix $C$. Moreover, these
transformations are diffeomorphisms and constitute a group.
}

A direct derivation of equation~(\ref{eq:10}) from the QP system~(\ref{eq:1})
is also possible~\cite{key-14} with the condition that the $n\times N$
matrix $A$ be of maximal rank~\cite{key-10}. In the case $N>n$, the trajectories of
the original $QP(A,B)$ system are, as in the case of the first canonical
form, restricted to invariants subspaces in the positive orthant of
the $N$-dimensional phase-space associated to the second canonical
form~(\ref{eq:10}). We stress here the
purely monomial dependence of the vector field of this canonical form.

\section{Urn processes and monomial differential systems}
\label{sec3}

In this chapter, we report the proof of a theorem obtained by
Flajolet, Dumas and Puyhaubert~\cite{key-1} mentioned in
the Introduction. We directly present the proof for $N$-colours urns
as these authors already gave it for two-colours urns processes
and assumed the proof for $N$ colours to be obvious to the reader.

Let us consider an urn with balls of $N$ possible colours. The process
is characterized by a $N\times N$ replacement matrix $M$ with integer
entries. We assume an initial composition vector $U^{0}=(u_{10},...,u_{N0})$
of the urn. A history of length $n$ of the urn process is a succession
of $n$ replacement steps of the urn's content starting from the initial
composition vector. It can be viewed as a trajectory of the urn's
content in the $N$-dimensional space of compositions. Since the urn
process is balanced, all histories of length $n$ are equiprobable.
Hence, the probability of finding in the urn
at step $n$ a composition vector $U=(u_{1},...,u_{N})$ is given
by the ratio of the number of histories of length $n$ starting at
$U^{0}$ and ending at $U$ over the total number of possible histories
of length $n$ starting at $U_{0}$. This statement can be written
as follows
\begin{equation}
P(U_{n}=U\mid U_{0}=U^{0})=\frac{\left[x_{1}^{u_{1}}....x_{N}^{u_{N}}z^{n}\right]
H(x_{1},...,x_{N},u_{10},...,u_{N0},z)}{\left[z^{n}\right]H(1,...,1,u_{10},...,u_{N0,}z)},
\label{eq:11}
\end{equation}
with the following definitions. The central tool here is
the counting generating function $H$ defined as
\begin{equation}
H(x_{1},...,x_{N},u_{10},...,u_{N0},z)\equiv\sum_{n:=0}^{\infty}\sum_{u_{1:}=0}^{\infty}...\sum_{u_{N}=0}^{\infty}
H_{n}(u_{10},...,u_{N0},u_{1},...,u_{N})x_{1}^{u_{1}}....x_{N}^{u_{N}}\frac{z^{n}}{n!},
\label{eq:12}
\end{equation}
where $H_{n}(u_{10},...,u_{N0},u_{1},...,u_{N})$ denotes
the number of histories that connect in $n$ steps the initial urn's
composition vector $U^{0}$ to an arbitrary composition vector $U$.
The notation $\left[x^{m}\right]S(x)$
represents as usual the coefficient of $x^{m}$ in the power series $S(x)$.

The association of a differential system with an urn process originates
from the analogy between, on one side, the actions of differentiation
and multiplication on monomials and, on the other, the replacement
procedures in an elementary step of the urn process. As simple examples,
consider the actions of the operators ${\partial}/{\partial x}$
and $x^{p+1}{\partial}/{\partial x}$ over the monomial $x^{m}$:
\begin{equation}
\frac{\partial}{\partial x}x^{m}=mx^{m-1}=x^{0}x...x+xx^{0}...x+...+xx...x^{0},
\label{eq:13}
\end{equation}
and 
\begin{equation}
x^{p+1}\frac{\partial}{\partial x}x^{m}=mx^{m+p}=(x^{1}x...x+xx^{1}...x+...+xx...x^{1})x^{p}.
\label{eq:14}
\end{equation}
The action~(\ref{eq:13}) can be interpreted in a combinatoric way as
choosing an object $x$ among $m$ similar objects in any possible
order and remove it. Whereas, equation~(\ref{eq:14}) corresponds
to choose an object $x$ among a set of $m$ similar objects in any
order, keep it in the set and add to it $p$ similar objects. 

Assuming that the variable $x_{i}$, $i\in\{1,...,N\},$ represents
balls of colour $i$, one can associate to a replacement matrix $M$
the partial differential operator
\begin{equation}
D\equiv\sum_{i=1}^{N}x_{1}^{M_{i1}}x_{2}^{M_{i2}}...x_{i}^{M_{ii}+1}...x_{N}^{M_{iN}}\frac{\partial}{\partial x_{i}},
\label{eq:15}
\end{equation}
In the same spirit, the monomial $x_{1}^{u_{1}}...x_{N}^{u_{N}}$
is associated to the composition vector of the urn $U$. Obviously,
the action of $D$ on that monomial
\begin{equation}
Dx_{1}^{u_{1}}...x_{N}^{u_{N}}=\sum_{i=1}^{N}u_{i}x_{1}^{u_{1}+M_{i1}}...x_{N}^{u_{N}+M_{iN}},
\label{eq:16}
\end{equation}
gives all possible transitions in one step of the urn's composition
that respect the prescriptions encoded in the replacement matrix $M$.
As a consequence, the $n$-iterated action of $D$ on the initial
composition monomial $x_{1}^{u_{10}}...x_{N}^{u_{N0}}$ generates
all the possible compositions of the urn after $n$ steps starting
from that initial composition. This can be expressed in terms of $H_{n}(u_{10},...,u_{N0},u_{1},...,u_{N})$,
the numbers of histories in $n$ steps linking the initial configuration
vector $U_{0}$ with a given final composition $U$, as follows:
\begin{equation}
D^{n}x_{1}^{u_{10}}...x_{N}^{u_{N0}}=\sum_{u_{1}=0}^{\infty}...\sum_{u_{N}=0}^{\infty}
H_{n}(u_{10},...,u_{N0},u_{1},...,u_{N})x_{1}^{u_{1}}....x_{N}^{u_{N}}.
\label{eq:17}
\end{equation}
From equations~(\ref{eq:12}) and~(\ref{eq:17}), we obtain the counting generating function:
\begin{equation}
H(x_{1},...,x_{N},u_{10},...,u_{N0},z)=\sum_{n=0}^{\infty}D^{n}x_{1}^{u_{10}}...x_{N}^{u_{N0}}\frac{z^{n}}{n!},
\label{eq:18}
\end{equation}
which can be summed as
\begin{equation}
H(x_{1},...,x_{N},u_{10},...,u_{N0},z)=e^{zD}x_{1}^{u_{10}}...x_{N}^{u_{N0}}.
\label{eq:19}
\end{equation}

With these results at hand, one can now prove the isomorphism found
by Flajolet and coworkers. Let us define the following system of ODEs:
\begin{equation}
\frac{dX_{i}}{dt}=X_{i}\prod_{j=1}^{N}X_{j}^{M_{ij}},
\label{eq:20}
\end{equation}
with $i\in\{1,...,N\}$ and where $t$ is a real variable. $M$ is
a constant matrix with integer entries and with balance $\sigma$.
As can be seen, the functions in the right-hand-side of this system
are monomials. The total degrees of each of these monomials are all
identical and equal to $\sum_{j=1}^{N}M_{ij}+1=\sigma+1$. 

Let us consider now any solution $X(t)=(X_{1}(t),...,X_{N}(t))$ of system~(\ref{eq:20}) and compute 
the derivative with respect to $t$ of the monomial
$X_{1}^{u_{1}}...X_{N}^{u_{N}}$ using equation~(\ref{eq:20}):
\begin{equation}
\frac{\partial}{\partial t}(X_{1}^{u_{1}}...X_{N}^{u_{N}})=\sum_{i=1}^{N}u_{i}X_{1}^{u_{1}+M_{i1}}...X_{N}^{u_{N}+M_{iN}}.
\label{eq:21}
\end{equation}
Obviously, this result correspond to the action~(\ref{eq:16}) of
the operator $D$ defined in~(\ref{eq:15})
\begin{equation}
\frac{\partial}{\partial t}(X_{1}^{u_{1}}...X_{N}^{u_{N}})=[Dx_{1}^{u_{1}}...x_{N}^{u_{N}}]_{\{x_{i}=X_{i};i=1,...,N\}},
\label{eq:22}
\end{equation}
and the $n$-th iteration of the operator $D$ gives 
\begin{equation}
\frac{\partial^{n}}{\partial t^{n}}(X_{1}^{u_{1}}...X_{N}^{u_{N}})=[D^{n}x_{1}^{u_{1}}...x_{N}^{u_{N}}]_{\{x_{i}=X_{i};i=1,...,N\}}.
\label{eq:23}
\end{equation}

As our aim is to make the connection between the differential system~(\ref{eq:20})
and the counting generating function $H$ of an urn process, let us
combine formula~(\ref{eq:18}) and~(\ref{eq:23}) to get
\begin{equation}
\sum_{n=0}^{\infty}\frac{z^{n}}{n!}\frac{\partial^{n}}{\partial t^{n}}[X_{1}^{u_{10}}(t)...X_{N}^{u_{N0}}(t)]=
X_{1}^{u_{10}}(t+z)...X_{N}^{u_{N0}}(t+z)=H(X_{1}(t),...,X_{N}(t),u_{10},...,u_{N0},z).
\label{eq:24}
\end{equation}
In the left hand side of the above equation, for $t$ and $z$ small
enough, the Taylor series can be applied thanks to the Cauchy-Kovalevskaya
theorem for the solutions of system~(\ref{eq:20}) which ensures
their existence and analyticity in the neighborhood of the origin.
Now, taking $t=0$ and z small enough for convergence in the above
formula, along with $(X_{1}(0)=x_{10},...,X_{N}(0)=x_{N0})$, we obtain:
\begin{equation}
H(x_{10},...,x_{N0},u_{10},...,u_{N0},z)=X_{1}^{u_{10}}(z)...X_{N}^{u_{N0}}(z).
\label{eq:25}
\end{equation}

We thus state the theorem:

\textbf{Theorem 2 }(Flajolet, Dumas, Puyhaubert) \textit{To any N-colour
balanced urn process of replacement matrix $M$ one can associate
one and only one differential system of the form $\dot{X}_{i}=X_{i}\prod_{j=1}^{N}X_{j}^{M_{ij}}$,
$j=1,...,N$. Moreover, the counting generating function of the histories
of the urn process with initial composition vector $(u_{10},...,u_{N0})$
is obtained in terms of the solution with initial condition $(X_{1}(0)=x_{10},...,X_{N}(0)=x_{N0})$
of the above differential system by the relation $H(x_{10},...,x_{N0},u_{10},...,u_{N0},z)=X_{1}^{u_{10}}(z)...X_{N}^{u_{N0}}(z)$.
}

Note that, in order to avoid divergences in case of negative entries
in $M$, the initial condition vector $(x_{10},...,x_{N0})$ of the
solution to the differential system must be in the open positive orthant.
The isomorphism between the monomial differential system~(\ref{eq:20})
and the urn process with the same matrix $M$ as replacement matrix
is, thus, established.

\section{General dynamical systems, urn processes and applications}
\label{sec4}

\subsection{Main result}

A simple glance at equations~(\ref{eq:10}) and~(\ref{eq:20}) shows
that they are strictly of the same form. The only restriction being
the fact that while the matrix $M$ can be quite general
in~(\ref{eq:10}), in the case of equation~(\ref{eq:20}) its entries
must be integers as they represent numbers of balls,
positive when they are introduced and negative when they are
extracted from the urn. Moreover, the matrix must be balanced. Nevertheless,
the formal identity between both differential systems and the use
of theorem 1 and theorem 2 allows us to state the following theorem:

\textbf{Theorem 3}\textit{ To any N-colour balanced urn process of
replacement matrix $M$ one can associate an infinite equivalence
class of quasi-polynomial systems of form $QP(A,B)$ given by equation
$\dot{x}_{i}=x_{i}\sum_{j=1}^{N}A_{ij}\prod_{k=1}^{n}x_{k}^{B_{jk}}$,
$i=1,...,n$, restricted to the open $N$-dimensional orthant
and with matrices $A$ and $B$ such that their matrix product $BA=M$.
This association is given by} $H(x_{10},...,x_{N0},u_{10},...,u_{N0},z)=X_{1}^{u_{10}}(z)...X_{N}^{u_{N0}}(z)$\textit{
and using the relations between the variables $X_{i}$, $i=1,...,N$, 
of the monomial canonical form and the variables $x_{i}$, $i=1,...,n$, 
of any $QP(A,B)$ system belonging to the equivalence class. }

 As reported above, a wide class of dynamical systems can be brought
to the QP form~\cite{key-9}. This permits to extend the above theorem
to that whole class of dynamical systems.

The above theorem paves the way to new techniques for studying both
urn processes and dynamical systems. As for finding solutions to an
urn process, one can try to solve any $QP(A,B)$ dynamical system
belonging to the equivalence class labeled by the replacement matrix
$M$ of this urn process. The solution of such a QP system can be
transformed into a solution of the monomial canonical form and, thence, 
by applying equation~(\ref{eq:25}) the generating function of the
urn process is obtained. Even though most QP systems cannot be exactly
solved due to their nonlinearity, many stability properties of their
solutions are known~\cite{key-15,key-17,key-18}.
This is due to the fact that being reducible to the Lotka-Volterra
canonical form, there is a systematic method to construct Lyapunov
functions for them. More generally, the monomial transformation~(\ref{eq:9})
connects the LV system to the monomial one and, furthermore, is a
diffeomorphism in the open positive orthant. One can, thus, transfer
all the mathematical properties that are preserved by that diffeomorphism
from the first to the the second canonical form. Among these properties,
chaotic regime have been systematically studied for the LV systems~\cite{key-19}
and should be of interest for the associated urn process. Another
result of interest for urn processes is the fact that
the solutions of the LV systems can be expressed in terms of a Taylor
series whose general coefficient is analytically known~\cite{key-20}.
These series define new special functions. All these results can be
transferred via transformation~(\ref{eq:9}) to the solutions of the
monomial differential system~(\ref{eq:20}) and, afterwards, to the
associated urn process via formula~(\ref{eq:25}).

As a corollary one can state:

\textbf{Corollary }\textsl{The solution to the N-colour balanced urn
process with replacement matrix $M$ and initial composition $\{u_{j0}=0,j\neq i;u_{i0}=1\}$
can be used to solve any $QP(A,B)$ system belonging to the equivalence
class labeled by the matrix $M$. This solution is obtained via the
relation
\begin{equation}
X_{i}(z)=H(x_{10},...,x_{N0},0,...,1,...0,z),
\label{eq:26}
\end{equation}
}
\textsl{with $i=1,...,N$.}

This provides the solution at time $z$ of the canonical system $\dot{X}_{i}=X_{i}\prod_{j=1}^{N}X_{j}^{M_{ij}}$,
$i=1,...,N$, with initial point $(x_{10},...,x_{N0})$. Next, using
the relation between the variables $X_{i}$, $i=1,...,N$, of the
monomial canonical system and the variables $x_{i}$, $i=1,...,n$,
of the original $QP(A,B)$ system, one gets the solution of the latter.

\subsection{Isomorphism between balanced urn processes and Lotka-Volterra dynamical
systems}

As an immediate consequence of theorems 1 and 2, we can state the
following result:

\textbf{Theorem 4 }\textsl{Any N-colour balanced urn process
of replacement matrix $M$ is isomorphic to a Lotka-Volterra system
of equations $\dot{Y}_{i}=Y_{i}\sum_{j=1}^{N}M_{ij}Y_{j}$, $i=1,...,N$,
provided $M$ is invertible.
}

\textbf{Proof: }The inverse of transformation~(\ref{eq:9}), written
here as $Y_{k}=\prod_{j=1}^{N}X_{j}^{M_{kj}}$, $k=1,...,N$, provided
the matrix $M$ is invertible, maps the monomial canonical form~(\ref{eq:20})
to the Lotka-Volterra form $\dot{Y}_{i}=Y_{i}\sum_{j=1}^{N}M_{ij}Y_{j}$,
$i=1,...,N$. This proves the proposition.

This theorem allows for transferring the abundant mathematical results
attached to the LV systems to the theory of urn processes~\cite{key-21}. 

\subsection{Example}

Consider the three-colours urn process with balance $\sigma=3$ and
replacement matrix
\begin{equation}
M=\left[\begin{array}{ccc}
1 & 0 & 2\\
1 & 1 & 1\\
2 & 1 & 0
\end{array}\right].
\label{eq:27-1}
\end{equation}

By theorem 2, the monomial differential system to which it is equivalent
is given by
\[
\dot{X_{1}}=X_{1}^{2}X_{3}^{2},
\]
\[
\dot{X_{2}}=X_{1}X_{2}^{2}X_{3},
\]
\begin{equation}
\dot{X_{3}}=X_{1}^{2}X_{2}X_{3}.
\label{eq:28-1}
\end{equation}
The above urn process is also equivalent, by theorem 4, to the following
Lotka-Volterra differential system
\[
\dot{X_{1}}=X_{1}(X_{1}+2X_{3}),
\]
\[
\dot{X_{2}}=X_{2}(X_{1}+X_{2}+X_{3}),
\]
\begin{equation}
\dot{X}_{3}=X_{3}(2X_{1}+X_{2}).
\label{eq:29-1}
\end{equation}

Moreover, theorem 3 tells us that this process is also equivalent to
an infinity of QP systems
\begin{equation}
\frac{dx_{i}}{dt}=x_{i}\sum_{j=1}^{3}A_{ij}\prod_{k=1}^{n}x_{k}^{B_{jk}},\qquad\mbox{ for }i=1,\ldots,n,
\label{eq:30-1}
\end{equation}
with $A$ and $B$ satisfying the constraint 
\begin{equation}
BA=M.
\label{eq:31-1}
\end{equation}

The natural number $n$ must satisfy the inequality $n\geq[{N}/{2}]$,
where $[r]$ for a real number $r$ is the smallest natural number
that is larger than $r$. In the present case of a 3-colours urn process we have $n\geq{2}$.
This inequality is due to the fact that the matrix equation~(\ref{eq:31-1})
corresponds to a system of $N^{2}$ multivariate quadratic polynomial
equations while the number of unknowns, i.e. the entries of the $n\times N$
matrix $A$ and those of the $N\times n$ matrix $B$, is $2nN$.
The above inequality ensures that this system is not overdetermined.

\subsection{Some consequences of theorems 2, 3 and 4}

\subsubsection{Equivalence between a urn process with matrix $M$
and a urn process with matrix $M-diag(M)$}

For that purpose, we have to introduce time reparametrizations. These transformations play an important
role in QP theory. The general form~(\ref{eq:1}) of a QP system
is covariant under a time reparametrization of the form:
\begin{equation}
dt=\prod_{i=1}^{n}x_{i}^{\beta_{i}}d\tau,\quad\beta_{i}\in\mathbb{R}.
\label{eq:27}
\end{equation}
Under such a transformation the equation~(\ref{eq:1}) becomes
\begin{equation}
\frac{d\tilde{x_{i}}}{d\tau}=\tilde{x}_{i}\sum_{j=1}^{N}\tilde{A}_{ij}
\prod_{k=1}^{n}\tilde{x}_{k}^{\tilde{B}_{jk}}\qquad\mbox{ for }i=1,\ldots,n\label{eq:28}
\end{equation}
with
\begin{equation}
\tilde{A}=A,
\label{eq:29}
\end{equation}
and 
\begin{equation}
\tilde{B_{jk}}=B_{jk}+\beta_{k},\qquad j=\mathrm{1},...,N;\:k=1,...,n.
\label{eq:30}
\end{equation}

Let us now apply this transformation with $\beta_{i}=-M_{ii}$,
$i=1,...,N$, to the monomial system~(\ref{eq:20}) associated to
the urn process with replacement matrix $M$ and balance $\sigma$.
This yields
\begin{equation}
\frac{dX_{i}}{d\tau}=X_{i}\prod_{j=1}^{N}X_{j}^{\tilde{M}_{ij}}\quad;i=1,...,N,
\label{eq:31}
\end{equation}
with $\tilde{M_{ij}}=M_{ij}-M_{jj}$ . This differential system is
associated to an urn process with replacement matrix $M-diag(M)$
and with balance $\tilde{\sigma}=\sigma-Tr(M)$. This allows us to
state the following theorem:

\textbf{Theorem 5} \textit{Any urn process with matrix $M$ and balance
$\sigma$ is isomorphic to a urn process with matrix $M-diag(M)$
and balance $\sigma-Tr(M)$}.

\subsubsection{Dimensional reduction of the $N$-dimensional
monomial differential system associated to a urn process with matrix
$M$ to a $N-1$ dimensional differential system}

\noindent \textbf{\ \ ~~~Theorem 6.}\textit{ Any monomial differential
system associated to a urn process with matrix $M$ and with balance $\sigma$:
\begin{equation}
\frac{du_{i}}{dt}=u_{i}\prod_{j=1}^{N}u_{j}^{M_{ij}}\:,\;\:i=1,\ldots,N,
\end{equation}
can be reduced to the $(N-1)$-dimensional QP system 
\begin{equation}
\frac{dy_{i}}{d\tau}=y_{i}\left(\prod_{j=1}^{N-1}y_{j}^{M_{ij}^{*}}-1\right)\:,\;\:i=1,\ldots,N-1,
\end{equation}
where $\tau$ arises from a time reparametrization,  $M_{ij}^{*}=M_{ij}-M_{Nj}$, $j=1,...,N$, and the $y_{j}$
are defined in terms of the $u_{j}$ by means of a monomial transformation.
}

\noindent \textbf{Proof}: We first perform the time reparametrization
$dt=\left(\prod_{i=1}^{N}u_{i}^{-M_{Ni}}\right)d{\tau}$ on the monomial
system associated to the urn process of matrix $M$. The outcome is
a system of the same form, now with exponent matrix: 
\begin{equation}
\tilde{M}=\left(\begin{array}{c}
M_{(N-1)\times N}^{*}\\
\hline \\
O_{1\times N}
\end{array}\right).
\end{equation}
where $M_{ij}^{*}=M_{ij}-M_{Nj}$, $j=1,...,N$, and where the indices
denote the sizes of each submatrix, while $O$ means the null matrix.
By construction, $\tilde{M}$ is a matrix of null balance, namely
\begin{equation}
\sum_{j=1}^{N}\tilde{M_{ij}}=0\:,\;\:i=1,\ldots,N.
\end{equation}
This zero-balance property leads naturally to the introduction of
the monomial transformation defined by the $N\times N$ regular matrix:
\begin{equation}
C=\left(\begin{array}{cccccc}
1 & 0 & 0 & \ldots & 0 & 1\\
0 & 1 & 0 & \ldots & 0 & 1\\
0 & 0 & 1 & \ldots & 0 & 1\\
\vdots & \vdots & \vdots & \ddots & 0 & 1\\
0 & 0 & 0 & \ldots & 1 & 1\\
0 & 0 & 0 & \ldots & 0 & 1
\end{array}\right).
\end{equation}
The result is a $QP(A^{*},B^{*})$ system with matrices 
\begin{equation}
B^{*}=B\cdot C=\tilde{M}\cdot C=\left(\begin{array}{ccc}
M_{(N-1)\times(N-1)}^{*} & \vline & O_{(N-1)\times1}\\
\hline  & \vline & \mbox{}\\
O_{1\times(N-1)} & \vline & 0
\end{array}\right).
\end{equation}
and 
\begin{equation}
A^{*}=C^{-1}\cdot A=C^{-1}=\left(\begin{array}{cccccc}
1 & 0 & 0 & \ldots & 0 & -1\\
0 & 1 & 0 & \ldots & 0 & -1\\
0 & 0 & 1 & \ldots & 0 & -1\\
\vdots & \vdots & \vdots & \ddots & 0 & -1\\
0 & 0 & 0 & \ldots & 1 & -1\\
0 & 0 & 0 & \ldots & 0 & 1
\end{array}\right).
\end{equation}
Accordingly, we are led to the reduced system 
\begin{equation}
\frac{dy_{i}}{d\tau}=y_{i}\left(\prod_{j=1}^{N-1}y_{j}^{M_{ij}^{*}}-1\right)\:,\;\:i=1,\ldots,N-1,
\end{equation}
and the trivial quadrature 
\begin{equation}
\frac{dy_{N}}{d\tau}=y_{N}.
\end{equation}
This proves the theorem.

The interpretation of the above result is that the constant balance
is actually a parametric constraint that can be exploited within the
operational framework of QP systems, leading to this dimensional reduction.

\section{Conclusions and perspectives}
\label{sec5}

The equivalence between urn processes and QP dynamical systems leads
to a wealth of consequences. The most evident among them are the transfer
of knowledge between the extensive literature on Lotka-Volterra systems
and the urn processes. As an example, one should exploit the fact that there exists 
a systematic method for constructing Lyapunov functions for fixed points of LV systems. Theorem
4, hence, allows us to apply these Lyapunov functions to the study
of the stability of fixed points for urn processes.
Conversely, results in balanced urn theory can be brought to the realm of LV  
and, more generally, QP systems. Urns for which the history counting generating function can be analytically found
may be used to find the solutions to the associated QP systems.\\

A more speculative application would be the numerical simulation
of dynamical systems via urn processes. Indeed, the computer simulation
of drawing balls at random from an urn and removing or adding others
could be quite fast. The probability of reaching a given composition
of the urn after a given number of steps should easily be computed from the
data and, in turn, the counting generating function $H$ could be obtained
by using relation~(\ref{eq:11}). The latter provides the solution
of the monomial canonical form~(\ref{eq:20}) through formula~(\ref{eq:26}).
From that solution, the solution of any QP systems belonging to the
same equivalence class could be computed. The difficulty, though, could
appear in relation with the calculation of the generating function
as the latter is given by an infinite series in term of the probability
(see equations~(\ref{eq:11}) and~(\ref{eq:12})).

To conclude, let us mention two open questions concerning the equivalence
between urn processes and differential systems. The first question
is the generalization of the above equivalence to the case of
unbalanced urn processes. In this case the fundamental relation~(\ref{eq:11})
looses its validity and the combinatorial approach adopted here can
no longer be used. The second question concerns the $QP(A,B)$ systems
whose matrices $A$ and $B$ have non-integer entries and are such
that their product $BA=M$ has non-integer entries. The question amounts
to the possibility of generalizing the urn processes, up to now characterized by
integer random variables, to processes that are similar but with real random variables.

\section{Acknowledgements}
I.G., A.~F. and T.~M.~R.~F.\ were partially financed by CNPq (Brazil).

\end{document}